\documentclass[a4paper]{jpconf}

\usepackage[english]{babel} 

\usepackage[%
	final,%
]{graphicx}

\usepackage[
   centertags, 
   sumlimits,  
   intlimits,  
   namelimits, 
]{amsmath} %
\usepackage{amssymb}

\usepackage[%
	pdftitle={Production, elliptic flow and energy loss of heavy quarks in the quark-gluon plasma},%
	pdfauthor={Jan Uphoff, Oliver Fochler, Zhe Xu, and Carsten Greiner},%
	pdfsubject={Production, elliptic flow and energy loss of heavy quarks at RHIC and LHC},
	pdfstartview=FitH,
	pdfpagemode=UseNone,
	bookmarksopen=true
	]{hyperref}

\usepackage[all]{hypcap} 

\usepackage{dcolumn} 

\usepackage{multirow} 

\usepackage{microtype} 

\makeatletter
\def\tagform@#1{\maketag@@@{\ignorespaces#1\unskip\@@italiccorr}}
\let\orgtheequation\theequation
\def\theequation{(\orgtheequation)}
\makeatother

\newcommand{\beq}{\begin{equation}}
\newcommand{\eeq}{\end{equation}}

\begin{document}
\title{Production, elliptic flow and energy loss of heavy quarks in the quark-gluon plasma}

\author{Jan Uphoff,$^1$ Oliver Fochler,$^1$ Zhe Xu,$^{2,1}$ and Carsten Greiner$^1$}

\address{$^1$ Institut f\"ur Theoretische Physik, Johann Wolfgang 
Goethe-Universit\"at Frankfurt, Max-von-Laue-Str. 1, 
D-60438 Frankfurt am Main, Germany}
\address{$^2$ Frankfurt Institute for Advanced Studies, Ruth-Moufang-Str. 1, D-60438 Frankfurt am Main, Germany}

\ead{uphoff@th.physik.uni-frankfurt.de}

\begin{abstract}
Production, elliptic flow and the nuclear modification factor of charm and bottom quarks are studied in central and non-central heavy-ion collisions at RHIC and LHC using the partonic transport model \emph{Boltzmann Approach of MultiParton Scatterings} (BAMPS). Employing an initial heavy quark yield obtained with PYTHIA the full space-time evolution of charm and bottom quarks in the quark-gluon plasma (QGP) is carried out with BAMPS, taking also secondary production in the QGP into account. Only elastic collisions of heavy quarks with particles from the medium cannot describe the experimentally observed elliptic flow and nuclear modification factor. However, using an improved Debye screening and the running coupling yields a result which is much closer to data.
\end{abstract}

\section{Introduction}
Due to their large mass heavy quarks are produced at an early stage of the collision and can cover -- depending on their production point -- a long distance through the QGP. Interaction along this path and subsequent modifications of the heavy quark can reveal valuable information about the properties of the medium.

The experimentally observed elliptic flow $v_2$  and nuclear modification factor $R_{AA}$  of heavy quarks \cite{Abelev:2006db,Adare:2006nq,Adare:2010de} hint at an energy loss which is comparable to that of light quarks. This is in contradiction with the expectations which one would draw from the ``dead cone effect'' \cite{Dokshitzer:2001zm,Zhang:2003wk}. Whether this large energy loss is due to collisional or radiative interactions -- or both (or even other effects) -- is under investigation 
(see \cite{Adare:2010de} for a recent overview and comparison with data).

After a short introduction to BAMPS we will briefly address the production of heavy quarks. In \autoref{sec:energy_loss} our results on the elliptic flow  and nuclear modification factor are discussed and compared to the experimental data.

\section{Parton cascade BAMPS}
\label{sec:bamps}
The partonic transport model \emph{Boltzmann Approach of MultiParton Scatterings} (BAMPS) \cite{Xu:2004mz,Xu:2007aa} simulates the full space-time evolution of the QGP by solving the Boltzmann equation,
\begin{equation}
\label{boltzmann}
\left ( \frac{\partial}{\partial t} + \frac{{\mathbf p}_i}{E_i}
\frac{\partial}{\partial {\mathbf r}} \right )\,
f_i({\mathbf r}, {\mathbf p}_i, t) = {\cal C}_i^{2\rightarrow 2} + {\cal C}_i^{2\leftrightarrow 3}+ \ldots  \ ,
\end{equation}
for on-shell partons and pQCD interactions. So far gluons (g) and heavy quarks (Q) are implemented with the following interactions: $g g \rightarrow g g $, $
        g g \rightarrow g g g $, $
        g g g \rightarrow g g    $, $
        g g \rightarrow Q  \bar{Q} $, $
        Q  \bar{Q} \rightarrow g g $, $
        g Q \rightarrow g Q $ and $
        g \bar{Q} \rightarrow g \bar{Q}$.
Details of the model and the employed cross sections can be found in \cite{Xu:2004mz,Xu:2007aa,Uphoff:2010sh}.

\section{Heavy quark production at RHIC and LHC}
We study the production of heavy quarks in two stages: initially during hard parton interactions in nucleon-nucleon scatterings and secondary production during the evolution of the QGP. Due to their large mass heavy quark production should be describable in the framework of pQCD. Indeed, next-to-leading order calculations \cite{Cacciari:2005rk} are in good agreement with experimental data from PHENIX \cite{Adare:2006hc_phenix_dsigmady}. For the present study, however, we use the event generator PYTHIA \cite{Sjostrand:2006za} to determine the initial heavy quark distributions, which agree with the PHENIX data as well.
Nevertheless, these distributions have large uncertainties due to their sensitivity on the parton distribution functions in nucleons, the heavy quark masses as well as the renormalization and factorization scales (see \cite{Uphoff:2010sh} for a detailed analysis).
For the initial distribution of the gluonic medium we use three different approaches: the mini-jet model, a color glass condensate inspired model and also PYTHIA in combination with the Glauber model.

In the following we  give a brief overview of our results on heavy quark production in the QGP. More details concerning this section can be found in \cite{Uphoff:2010sh}.

Secondary heavy quark production in the QGP is studied within a full BAMPS simulation of Au+Au collisions at RHIC.
According to our calculations the charm quark production in the medium lies between 0.3 and 3.4 charm pairs, depending on the model of the initial gluon distribution, the charm mass and whether a $K=2$ factor for higher order corrections of the cross section is employed. However, compared to the initial yield these values are of the order of a few percent for the most probable scenarios. Therefore, one can conclude that charm production at RHIC in the QGP is nearly negligible. 

At LHC, however, the picture looks a bit different: Here the charm production in the QGP is a sizeable fraction of the initial yield and is even of the same order for some scenarios (with mini-jet initial conditions for gluons with a high energy density). In numbers, between 11 and 55 charm pairs are produced in the QGP.

Bottom production in the QGP, however is very small both at RHIC and LHC and can be safely neglected. As a consequence, all bottom quarks at these colliders are produced in initial hard parton scatterings.

\section{Elliptic flow and nuclear modification factor of heavy quarks at RHIC}
\label{sec:energy_loss}
Commonly used observables for investigating the coupling of heavy quarks with the medium are the elliptic flow
\begin{align}
\label{elliptic_flow}
	v_2=\left\langle  \frac{p_x^2 -p_y^2}{p_T^2}\right\rangle 
\end{align} 
($p_x$ and $p_y$ are the momenta in $x$ and $y$ direction in respect to the reaction plane)
and the nuclear modification factor
\begin{equation} \label{eq_hard_probes_RAA}
R_{AA}=\frac{{\rm d}^{2}N_{AA}/{\rm d}p_{T}{\rm d}y}{N_{\rm bin} \, {\rm d}^{2}N_{pp}/{\rm d}p_{T}{\rm d}y}
\end{equation}
of heavy quarks at mid-rapidity. A large elliptic flow comparable to that of light partons indicates a strong coupling to the medium. On the other hand a small $R_{AA}$ is a sign for a large energy loss of heavy quarks. Experimental results reveal that both quantities are on the same order as the respective values for light particles \cite{Abelev:2006db,Adare:2006nq,Adare:2010de}.

As we have recently shown \cite{Uphoff:2010fz} elastic scatterings of heavy quarks with the gluonic medium using a constant coupling $\alpha_s = 0.3$ and the Debye mass for screening the $t$ channel cannot reproduce the experimentally measured elliptic flow.  In order to explain the data one would need a $40-50$ times larger cross section than the leading order one. Of course, this $K$ factor is too large to represent the contribution of higher order corrections. However, as we will show in this section, the discrepancy with the data can be lowered -- even on the leading order level -- by a factor of 10 by taking the running of the coupling into account and by improving the Debye screening. The remaining factor of 4 difference could then indeed stem from neglecting higher order effects, which, however, must be checked in a future project.

The following calculations are done analogously to \cite{Gossiaux:2008jv,Peshier:2008bg}. An effective running coupling is obtained from measurements of $e^+e^-$ annihilation and non-strange hadronic decays of $\tau$ leptons \cite{Dokshitzer:1995qm,Gossiaux:2008jv}: 
\begin{align}
 \alpha_s(Q^2)= \frac{4\pi}{\beta_0} \begin{cases}
  L_-^{-1}  & Q^2 < 0\\
  \frac12 - \pi^{-1} {\rm atn}( L_+/\pi ) &  Q^2 > 0
\end{cases}
\end{align}
with $\beta_0 = 11-\frac23\, N_f$, $N_f=0$ in our case, and
$L_\pm = \ln(\pm Q^2/\Lambda^2)$ with $\Lambda=200 \, {\rm MeV}$. If $\alpha_s(Q^2)$ is larger than $\alpha_s^{\rm max} = 1.0$ it is set to $\alpha_s^{\rm max}$.

Since the $t$ channel of the $g Q \rightarrow g Q$ cross section is divergent it is screened with a mass proportional to the Debye mass $m_{D}$:
\begin{align}
\label{t_screening}
	 \frac{1}{t} \rightarrow \frac{1}{t-\kappa \, m_{D}^2}
\end{align}
The Debye mass is calculated by the common definition $m_{D}^2 = 4 \pi \, (1+N_f/6) \, \alpha_s(t) \, T^2$ with the running coupling. The prefactor $\kappa$ in \autoref{t_screening} is mostly set to 1 in the literature without a clear reason. However, one can fix this factor by comparing the ${\rm d}E/{\rm d}x$ of the born cross section with $\kappa$ to the energy loss within the hard thermal loop approach to $\kappa \approx 0.2$ \cite{Gossiaux:2008jv,Peshier:2008bg}.

\begin{figure}
\begin{minipage}[t]{0.49\textwidth}
\centering
\includegraphics[width=1.0\textwidth]{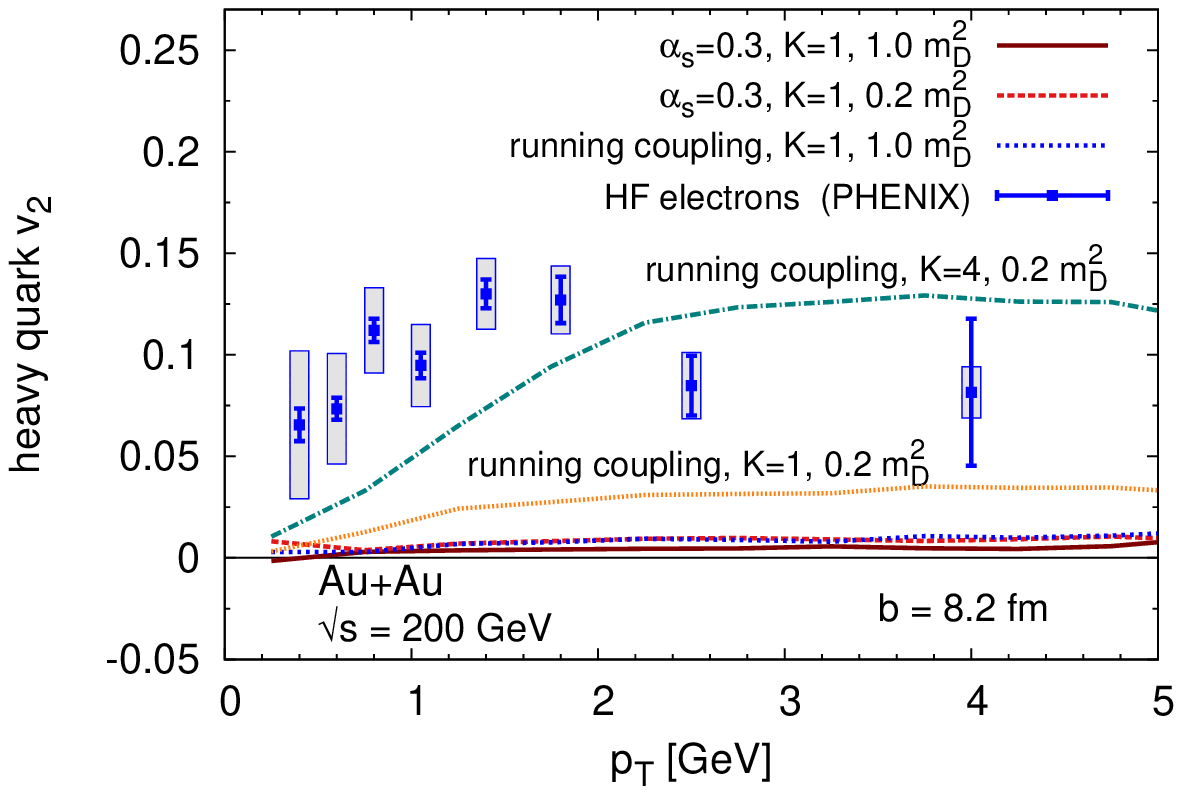}
\end{minipage}
\hfill
\begin{minipage}[t]{0.49\textwidth}
\centering
\includegraphics[width=1.0\textwidth]{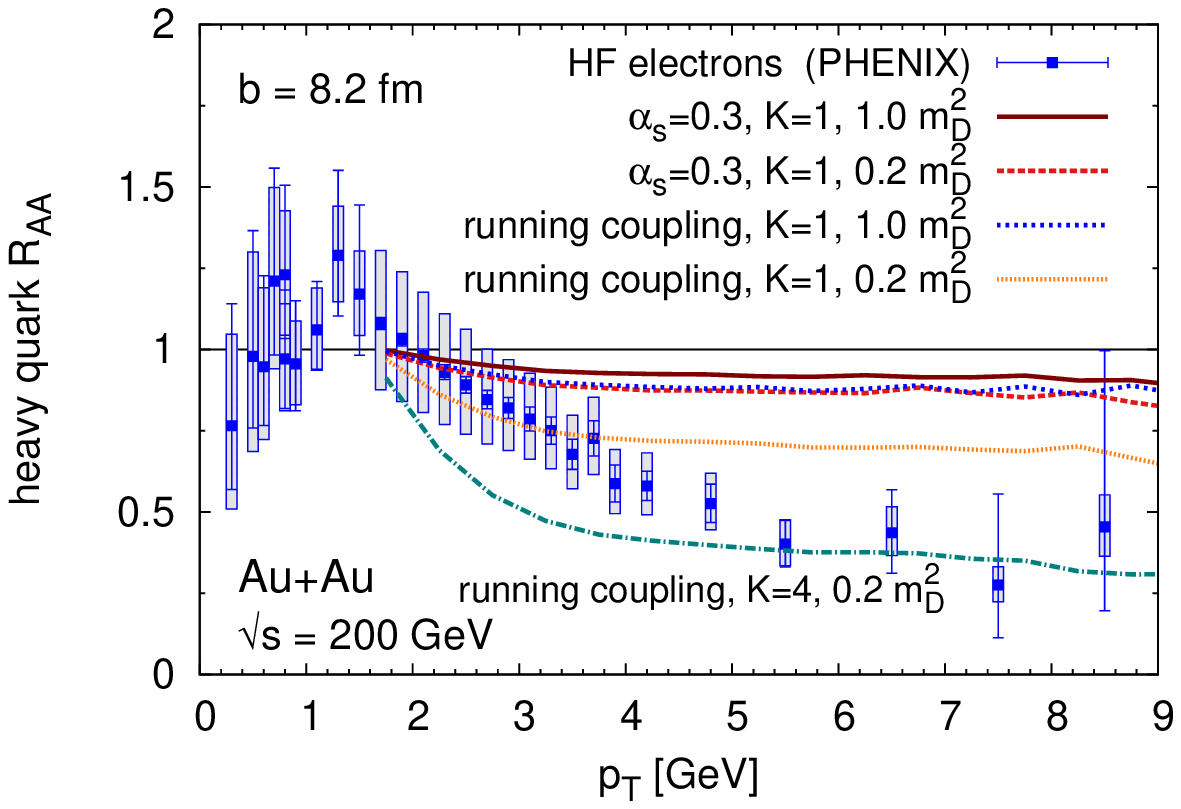}
\end{minipage}
\caption{Elliptic flow $v_2$ (left) and nuclear modification factor $R_{AA}$ (right) of heavy quarks with pseudo-rapidity $|\eta|<0.35$ at the end of the QGP phase for Au+Au collisions at RHIC with an impact parameter of $b=8.2 \, {\rm fm}$. For one curve the cross section of $gQ \rightarrow gQ$ is multiplied with a $K$ factor. For comparison data of heavy flavor electrons \cite{Adare:2010de} is shown (see also text).}
\label{fig:v2_raa}
\end{figure}

These improvements lead to an enhanced cross section which also increases the elliptic flow. \autoref{fig:v2_raa} shows $v_2$ as a function of the transverse momentum $p_T$ for the leading order cross section without any improvements, with the running coupling, with the corrected Debye screening and with both modifications.
The elliptic flow of the latter reproduces the order of magnitude of the data, if the cross section is multiplied with $K=4$, which is much smaller than the previous employed $K=40-50$ and lies in a region which could account for higher order corrections. However, one has to check if these corrections have indeed a similar effect as a constant $K$ factor of 4. Therefore, the calculation of the next-to-leading order cross section is planned for the near future and will complement $2 \leftrightarrow 3$ interactions for gluons, which are already implemented in BAMPS \cite{Xu:2004mz}. The shapes of the theoretical curve and of the data points are, however, slightly different. This is probably an effect of hadronization and decay to electrons, which is neglected in this calculation. That is, in \autoref{fig:v2_raa} $v_2$ and $R_{AA}$ of heavy quarks on the quark level are compared to the heavy flavor electron data. We plan to take these effects into account in future investigations.

On the right hand side in \autoref{fig:v2_raa} the $R_{AA}$ of heavy quarks is depicted, which shows for $K=4$ the same magnitude of suppression as the data. Studies on the $R_{AA}$ of gluons in BAMPS are presented in \cite{Fochler:2008ts,Fochler:2010wn}.

\section{Conclusions}

We have studied the production and space-time evolution of charm and bottom quarks with BAMPS. Charm production in the QGP at RHIC is to a good approximation negligible, but at LHC it can reach values comparable to the initial yield. Bottom production during the evolution of the medium can be neglected at RHIC and LHC.

The energy loss of heavy quarks in the QGP due to elastic collisions with the leading order cross section is very small. Nevertheless, if one improves the calculation of the cross section by taking the running coupling and a more precise Debye screening into account, the order of magnitude of the experimentally measured elliptic flow and nuclear modification factor can be reproduced, if one multiplies the cross section with a $K$ factor of 4. In order to solve this discrepancy next-to-leading order contributions must be taken into account, which we plan to do in the future.

\section*{Acknowledgements}
We would like to thank A. Peshier and P.B. Gossiaux for stimulating and helpful discussions.

The BAMPS simulations were performed at the Center for Scientific Computing of the Goethe University Frankfurt. This work was supported by the Helmholtz International Center for FAIR within the framework of the LOEWE program launched by the State of Hesse.

\section*{References}
\bibliography{hq}

\end{document}